\documentclass[vecphys]{svmult}

\usepackage{makeidx}         % allows index generation
\usepackage{graphicx}        % standard LaTeX graphics tool
\usepackage{multicol}        % used for the two-column index
\usepackage[bottom]{footmisc}% places footnotes at page bottom
\usepackage{subfigure}
\newcommand{\Lya}{Ly-$\alpha$~}

\makeindex             % used for the subject index
\begin{document}

\title*{The contribution of UVES@VLT to the new era of QSO
  absorption line studies} 
\titlerunning{UVES@VLT and QSO absorption line studies} 
% for an abbreviated version of
\author{Valentina D'Odorico\inst{1}\and
Miroslava Dessauges-Zavadsky\inst{2}}
\institute{INAF - Osservatorio Astronomico di Trieste, via G.B. Tiepolo
  11, 34143-I Trieste, Italy
\texttt{dodorico@oats.inaf.it}
\and Observatoire Astronomique de l'Universit\'e de Gen\`eve, ch. des
  Maillettes 51, 1290 Sauverny, Switzerland
  \texttt{miroslava.dessauges@obs.unige.ch}}  
\maketitle

\begin{abstract}
We briefly review the main results obtained in the field of QSO
absorption line studies with the UVES high
resolution spectrograph  mounted on the Kueyen unit 
of the ESO Very Large Telescope (Paranal, Chile).  
%We apologize for those results that were not mentioned due to the limited space.  
\end{abstract}

\section{Introduction}
\label{sec:1}

Over the past decade, our understanding of the intergalactic medium
(IGM) at high redshift, $z=2-5$, the  main baryonic component of the
cosmic web, has advanced considerably.  
%Qualitative questions concerning the nature and interpretation of the
%IGM have given way increasingly %to quantitative investigations aimed
%at measuring properties of the diffuse baryons, among them the
%%temperature, metallicity, kinematics and radiation field.   
We are now able to measure properties of the diffuse baryons,  among
them the temperature, metallicity, kinematics and radiation field, and
obtain their distributions  as functions of time, spatial scale, and
density.  
These improvements are primarily determined by a rich amount of new
high-resolution observational data (mainly from UVES at the VLT and
HIRES at Keck) and by new theoretical hydrodynamical simulations, that
incorporate the relevant physical processes. These two factors have
determined a paradigm shift in the study of absorption line systems:
they are now considered as tracers of the entire cosmic structure
formation process over the cosmic history, and not simple probes of
physical processes taking place at the Jeans length (see
\cite{meiksin} for a recent review).  

\section{Why UVES made the difference?}

\begin{figure}
\centering
\includegraphics[height=4cm]{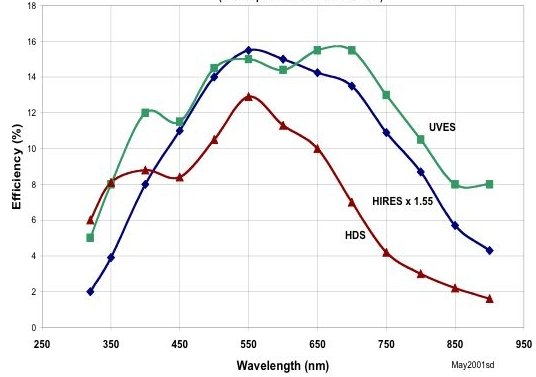}
\caption{Efficiency curves for UVES, HIRES (before August 2004), and HDS
  including telescope, instrument and detector as a function of
  wavelength. Courtesy of S. D'Odorico. }
\label{fig:UVES}       % Give a unique label
\end{figure}
The UV-Visual Echelle Spectrograph \cite{dekker} is the
high-resolution ($R\sim 40,000$ with 1-arcsec slit) optical
spectrograph of the ESO VLT. It started operation in fall 1999.  
The  high efficiency from the atmospheric
cut-off at 300 nm to the long wavelength limit of the CCD detectors
(about 1100 nm) was determinant for UVES to excel among the
equivalent instruments (HIRES at Keck and HDS at Subaru, see
Fig.~\ref{fig:UVES}). 
In our view, two other features contributed to the excellent 
scientific results obtained with UVES: the dedicated data reduction
pipeline \cite{ballester}, working since the beginning of operations, and 
the availability of observed data in the public archive of ESO  
which, in the specific case of the data rich QSO spectra, allowed different users 
to `squeeze' all the possible science out of them.

\section{The IGM probed by single lines of sight}

A remarkable contribution to the study of the IGM with the \Lya forest was given by 
the ESO Large Programme (LP) ``Cosmic Evolution of the IGM''
\cite{bergeron}.  
%A total of 334 h of UVES exposure time (with seeing $\le 0.8$ arcsec) 
%was dedicated to this programme, resulted in a 
A sample of 19 QSO spectra at $R\sim 45,000$ with $S/N\sim35$ and 70
per pixel at 350 and 600 nm, respectively, was collected covering the
\Lya forest between $z\sim1.7$ and 3.5.   
Data were immediately released to the public and, as of August
2007, they were used in 29 refereed publications generating 658
citations in 3 years\footnote{Source: ``Telescope Bibliography'' maintained by the
ESO Library}.  

\begin{figure}
\centering
\includegraphics[height=6cm]{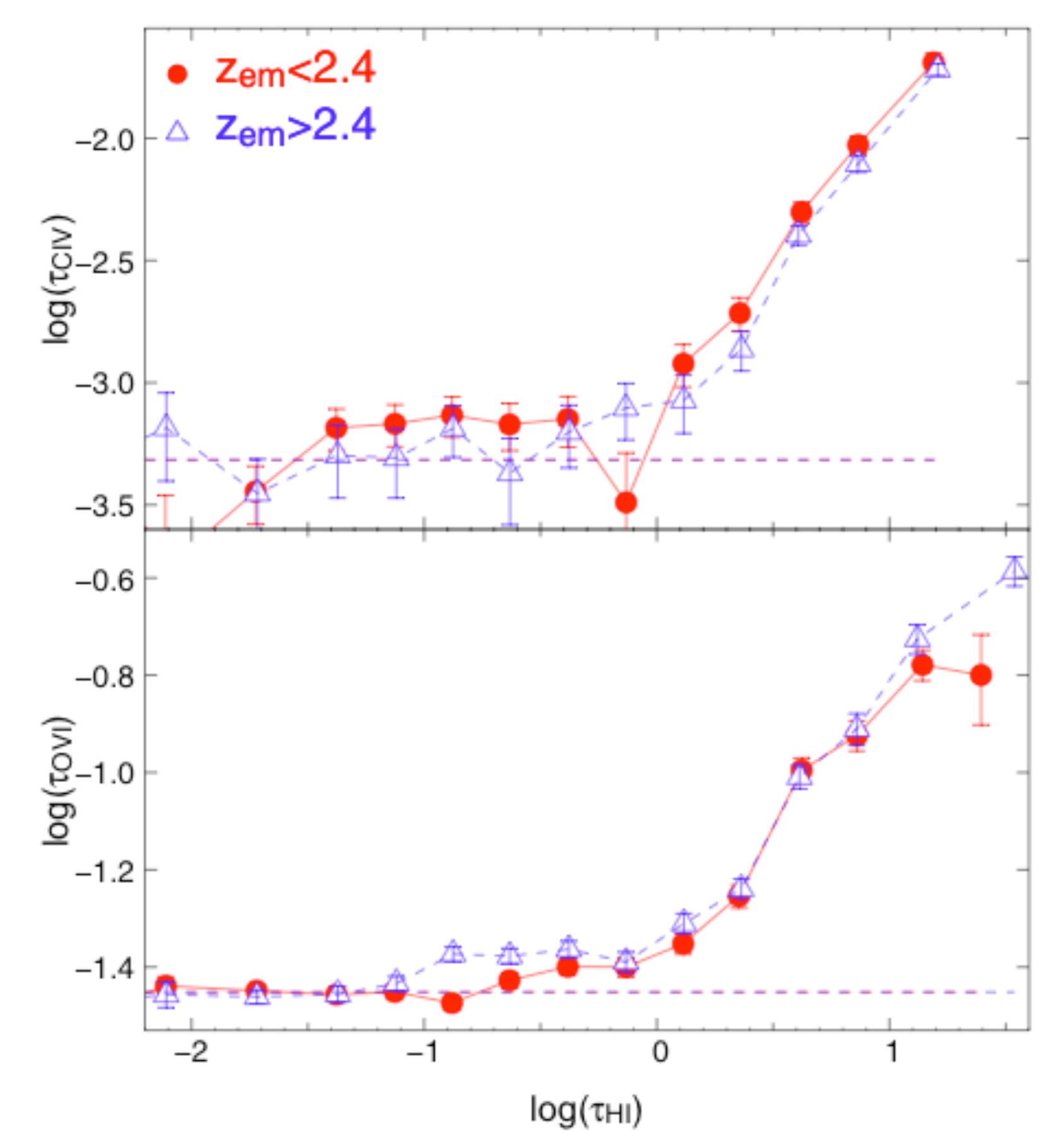}
\label{fig:metal}
\caption{C~IV (top panel) and O~VI (bottom panel) optical depth
  plotted versus the H~I optical depth for the LP QSO sample (from \cite{aracil04})}
\end{figure}

%The big effort made to collect this exceptional sample, was rewarded by
%the great relevance of the obtained results. 
The following important results were obtained using this sample. 
The investigation of the metal content of the low density IGM with the
``pixel optical depth statistics'' (see details in \cite{aguirre}), to
discriminate between early and late enrichment scenarios.  
Metal ions are detected down to the mean cosmic density, but the
information in the under-dense regions, which is critical for the
studied issue, is still poor (see Fig.~\ref{fig:metal} and
\cite{aracil04}).  
The nature of the \Lya forest was exploited to compute  the
transmitted flux \cite{kim04} and the dark matter power spectra,
allowing to tighten the values of the cosmological parameters derived
from the CMB \cite{viel04}.  
The statistical and physical properties of \Lya
(e.g. \cite{kim02,saitta07}) and C~IV (e.g. \cite{scanna}) absorbers
were refined and assessed.  

Another topic that was extensively investigated with UVES spectra was
the time-variability of fundamental constants (see the contribution by
P. Molaro to these proceedings).

\begin{figure}
\centering
\includegraphics[height=5cm]{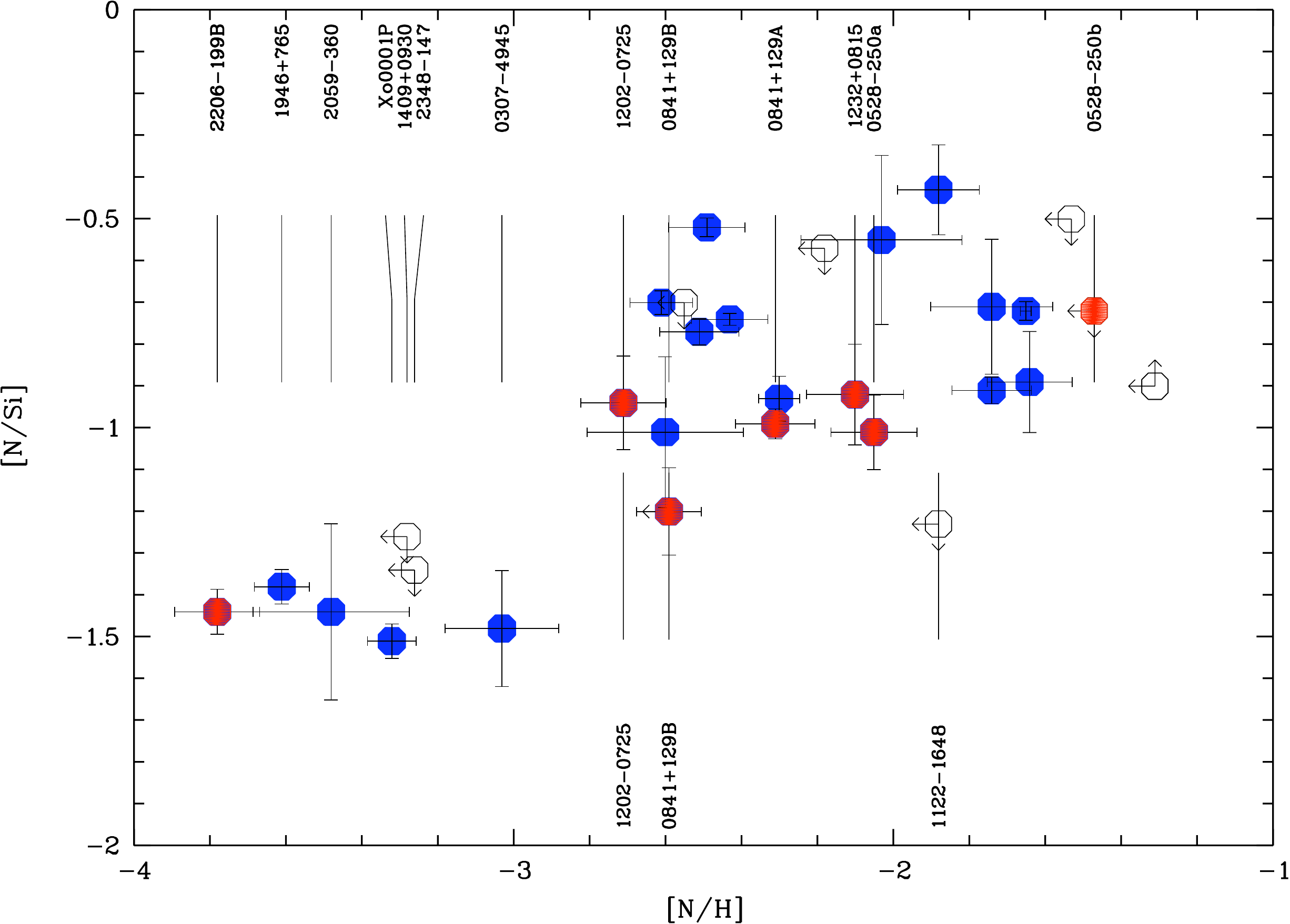}
\label{fig:nitrogen}
\caption{[N/$\alpha$] ratio  as a function of Nitrogen abundance for
  DLA absorbers (from \cite{dodomolaro})} 
\end{figure}

\section{High redshift galaxies traced by Damped \Lya systems}

Damped \Lya systems (DLA) are the quasar absorption line systems  with
the highest H~I column densities ($N$(HI) $> 2\times 10^{20}$
cm$^{-2}$) and  they arise in galactic disks or halos. They are
invaluable tools to study the chemical abundances in the interstellar
medium of objects in the very young Universe.  

At present, there are more than 60 QSO spectra with a DLA available in
the ESO public archive.  Some of the notable results obtained in this
field are: the estimate of the temperature of the CMB radiation at
$z>2$, which is in agreement with the hot Big Bang cosmology predictions
\cite{srianand00,molaro02}; the hints on the nucleosynthesis of 
Nitrogen in young objects (see Fig.~\ref{fig:nitrogen},
\cite{miriam03}), the properties of molecular hydrogen in DLA
\cite{ledoux03,srianand05,petitjean06}, and the star
formation histories of individual DLA absorbers \cite{mirka04,mirka07}.  

The contribution of the sub-DLA population  ($N({\rm HI} \ge 10^{19}$
cm$^{-2}$) to the total HI gas mass and to the missing metal problem
was also investigated \cite{celine05,celine07}.

\begin{figure}
\centering
\includegraphics[height=5cm]{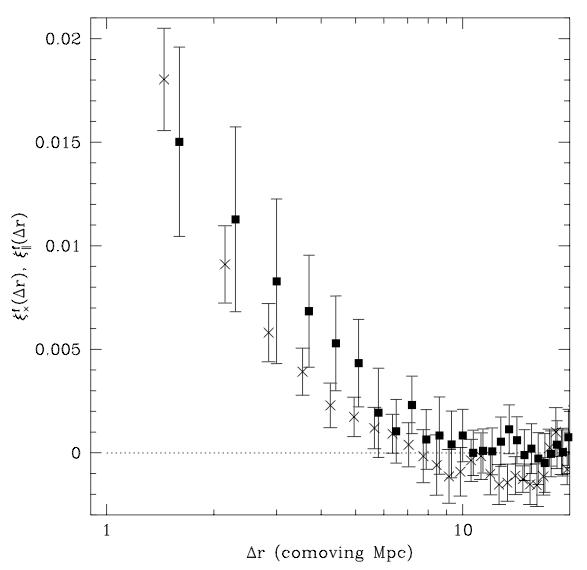}
\caption{Comparison of the cross-correlation function for the sample of
QSO pairs (squares) with the auto-correlation function computed for the LP
QSO sample (crosses) as a function of comoving spatial separation across
and along the line of sight, respectively. The cross-correlation function is
slightly shifted in $\Delta r$ for clarity (from \cite{vale06}).}
\label{fig:pair}       % Give a unique label
\end{figure}

\section{The tomography of the IGM}

%Studies of the IGM along single lines of sight are lacking the
%information in the transverse direction.    
To reconstruct the 3D distribution of diffuse matter with its physical
and chemical properties and investigate the small-scale IGM-galaxy
interactions, many high-redshift bright sources, close in the sky, are
needed to use as background lights for spectroscopic studies.  
Several QSO pairs and groups were identified by the recent large
surveys (2dFQRS, SDSS), however, most of them are too faint to be
observed with UVES ($V>18$).    

A great effort was put in collecting a statistically significant
sample of UVES spectra of close QSO pairs that was used to investigate
the transverse clustering properties of strong metal absorbers
\cite{vale02} and of the transmitted flux in the \Lya forest
\cite{vale06}.  
The agreement of the transmitted flux correlation functions along and
across the line of sight  (see Fig.~\ref{fig:pair}) implied that
distortions in redshift space 
due to peculiar velocities are small ($ < 100$ km s$^{-1}$) and
confirms the validity of the concordance cosmological model. 

\section{Future perspectives}

%We foresee three decisive steps for the development of QSO absorption
%line studies in the next future. 
In the second half of 2008, first light is planned for  X-shooter at the VLT  
\cite{vernet}: a single target, intermediate resolution (5000 at UVB
and NIR, 7000 at VIS for 1 arcsec slit) wide wavelength range (UV to K
bands) spectrograph.  This instrument was conceived to study QSO pairs
and groups and will finally allow to carry out  the Alcock-Paczy\`nski
test, sensible to the cosmological energy content (e.g. \cite{aptest}).   
For the next decade, two new spectrographs are under study:  CODEX at
the ELT (see the contribution by J. Liske to these proceedings) and
its precursor at the VLT, ESPRESSO (see the contribution by
L. Pasquini to these proceedings). They will be characterised by a
very high resolution ($R\sim 150,000$), high efficiency, high
stability and new design concepts.   A second revolution in QSO
absorption line studies is approaching fast! 

% Non-BibTeX users please follow the syntax
% the syntax of "referenc.tex" for your own citations
%%%%%%%%%%%%%%%%%%%%%%%% referenc.tex %%%%%%%%%%%%%%%%%%%%%%%%%%%%%%
% sample references
% "physics"
%
% Use this file as a template for your own input.
%
%%%%%%%%%%%%%%%%%%%%%%%% Springer-Verlag %%%%%%%%%%%%%%%%%%%%%%%%%%

%
% BibTeX users please use
% \bibliographystyle{}
% \bibliography{}

%%%%%%%%%%%%%%%%%%%%%%%%%%%%%%%%%%%%%%%%%%%%%%%%%%%%%%%%%%%%%%%%%%%%%%  }

%%%%%%%%%%%%%%%%%%%%%%%%%%%%%%%%%%%%%%%%%%%%%%%%%%%%%%%%%%%%%%%%%%%%%%

\printindex
\end{document}